\documentclass[manuscript]{acmart}
\AtBeginDocument{%
  \providecommand\BibTeX{{%
    \normalfont B\kern-0.5em{\scshape i\kern-0.25em b}\kern-0.8em\TeX}}}

\setcopyright{acmcopyright}
\copyrightyear{2021}
\acmYear{2021}
\acmDOI{10.1145/1122445.1122456}

\acmConference[UIST '21]{ACM Symposium on User Interface Software and Technology}{October 10--13, 2021}{virtual}
\acmBooktitle{ACM Symposium on User Interface Software and Technology (UIST '21), October 10--13, 2021, virtual}
\acmPrice{15.00}
\acmISBN{978-1-4503-XXXX-X/18/06}

\acmSubmissionID{8737}

\usepackage{enumitem}
\usepackage{amsmath}
\usepackage{amsfonts}
\usepackage{multirow}
\usepackage{color}
\usepackage{booktabs}
\usepackage{ifthen}
\usepackage{hyperref}
\usepackage{subcaption}
\usepackage[most]{tcolorbox}
\usepackage{soul}
\usepackage{cuted}
\usepackage{capt-of}

\usepackage{xcolor}
\usepackage[normalem]{ulem} 

\usepackage{algorithm}
\usepackage{algorithmicx}
\usepackage{algpseudocode}
\algnewcommand\algorithmicinput{\textbf{Input:}}
\algnewcommand\INPUT{\item[\algorithmicinput]}
\algnewcommand\algorithmicoutput{\textbf{Output:}}
\algnewcommand\OUTPUT{\item[\algorithmicoutput]}
\algnewcommand\algorithmicforeach{\textbf{for each}}

\algdef{S}[FOR]{ForEach}[1]{\algorithmicforeach\ #1\ \algorithmicdo}
\algrenewcommand{\alglinenumber}[1]{\color{red!80!blue}\footnotesize#1:}

\algnewcommand\Func[2]{\textcolor{green}{#1}\textcolor{green}{(#2)}}
\algnewcommand\Insert[2]{Insert {#1} to #2.}
\algnewcommand\Input[1]{\State \textbf{Input: } #1}
\algnewcommand\Output[1]{\State \textbf{Output: } #1}

\definecolor{gray}{rgb}{0.5,0.5,0.5}
\definecolor{green}{rgb}{0, 0.6, 0}
\definecolor{orange}{rgb}{1, 0.5, 0}
\definecolor{mahogany}{rgb}{0.75, 0.25, 0.0}
\definecolor{purple}{rgb}{0.6, 0, 0.6}
\definecolor{darkgreen}{rgb}{0, 0.3, 0}
\definecolor{orange}{rgb}{1, 0.5, 0.}

\newcommand{\myhl}[1]{{#1}}
\newcommand{\chl}[1]{#1}

\newcommand{\ignore}[1]{}
\newcommand{\none}[1]{}
\newcommand{\com}[1]{}

\newcommand{\ie}{i.e.,}
\newcommand{\eg}{e.g.,}
\newcommand{\figname}{Figure}

\newcommand{\secname}{Section}

\begin{document}

\title{Per Garment Capture and Synthesis for Real-time Virtual Try-on}

\author{Toby Chong}
\email{tobyclh@gmail.com}
\orcid{1234-5678-9012}
\affiliation{%
  \institution{The University of Tokyo, Hongo}
  \city{Tokyo}
  \country{Japan}
}
\author{I-Chao Shen}
\email{ichao.shen@ui.is.s.u-tokyo.ac.jp}
\affiliation{%
  \institution{The University of Tokyo, Hongo}
  \city{Tokyo}
  \country{Japan}
}
\author{Nobuyuki Umetani}
\email{umetani@ci.i.u-tokyo.ac.jp}
\affiliation{%
  \institution{The University of Tokyo, Hongo}
  \city{Tokyo}
  \country{Japan}
}
\author{Takeo Igarashi}
\email{takeo@acm.org}
\affiliation{%
  \institution{The University of Tokyo, Hongo}
  \city{Tokyo}
  \country{Japan}
}


\begin{abstract}
Virtual try-on is a promising application of computer graphics and human computer interaction that can have a profound real-world impact especially during this pandemic.
Existing image-based works try to synthesize a try-on image from a single image of a target garment, but it inherently limits the ability to react to possible interactions.
It is difficult to reproduce the change of
wrinkles caused by pose and body size change, as well as pulling and stretching of the garment by hand.
In this paper, we propose an alternative per garment capture and synthesis workflow to handle such rich interactions by training the model with many systematically captured images.
Our workflow is composed of two parts: garment capturing and clothed person image synthesis.
We designed an actuated mannequin and an efficient capturing process that collects the detailed deformations of the target garments under diverse body sizes and poses.
Furthermore, we proposed to use a custom-designed measurement garment, and we captured paired images of the measurement garment and the target garments.
We then learn a mapping between the measurement garment and the target garments using deep image-to-image translation.
The customer can then try on the target garments interactively during online shopping.
The proposed workflow requires certain manual labor, but we believe that the cost is acceptable given that the retailers are already paying significant costs for hiring professional photographers and models, stylists, and editors to take photographs for promotion.
Our method can remove the need of hiring these costly professionals.
We evaluated the effectiveness of the proposed system with ablation studies and quality comparison with previous virtual try-on methods.
We perform a user study to show our promising virtual try-on performances.
Moreover, we also demonstrate that we use our method for changing virtual costumes in video conferences.
Finally, we provide the collected dataset as the cloth dataset parameterized by various viewing angles, body poses, and sizes\footnote{Project page: https://sites.google.com/view/deepmannequin/home}.


\end{abstract}



\begin{teaserfigure}
  \includegraphics[width=\textwidth]{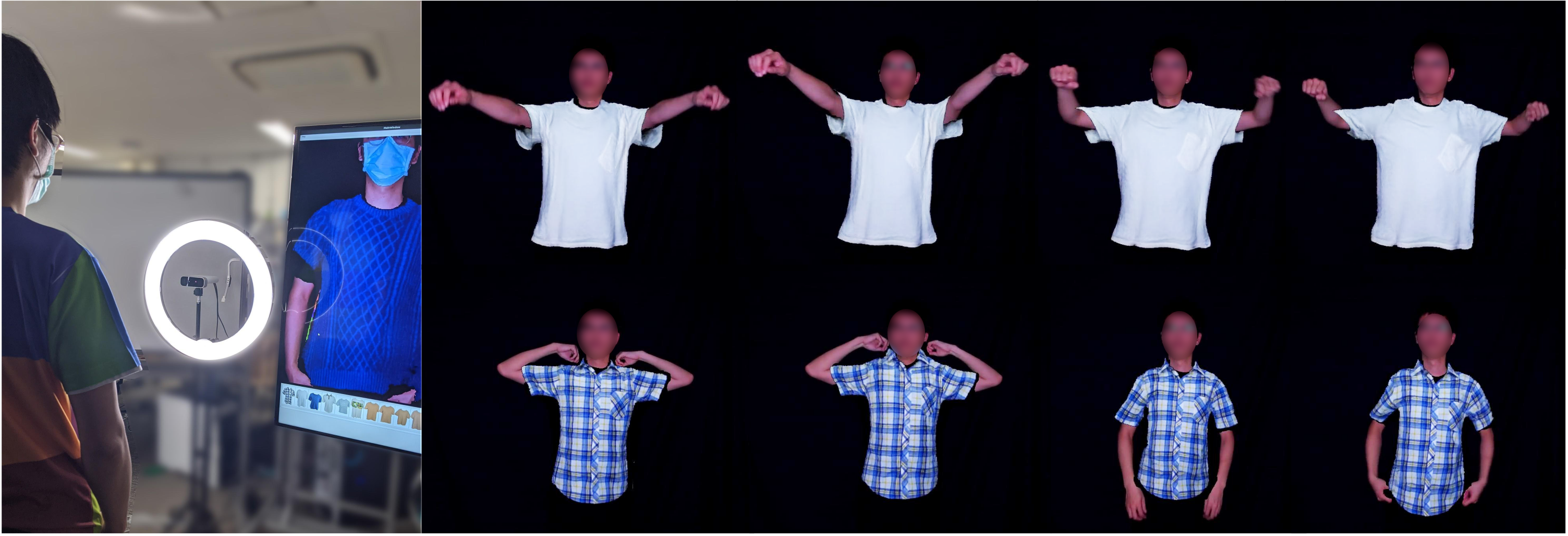}
  \caption{Our real-time system allows the user to virtually try-on various garments using different interactions, such as stretching the body and pulling the garment.
With the proposed per garment and synthesis workflow, the detailed wrinkles are captured and synthesized according to different motions.}
  \Description{Enjoying the baseball game from the third-base
  seats. Ichiro Suzuki preparing to bat.}
  \label{fig:teaser}
\end{teaserfigure}

\maketitle

\section{Introduction}
The emergence of online shopping significantly advances the efficiency and convenience of shopping for both the customers and the retailers.
Especially during the current pandemic, people often choose to shop online for various goods, including daily groceries, electrical products, and fashion apparel instead of store visit.
Among all the types of the good we can shop online, fashion apparels require the most subjective decision thus poses the most significant challenge for online shopping.
More specifically, the experience of how a garment fits the customer himself plays the most critical role in the buying decision.
However, most of the existing online shopping websites fail to provide satisfying in-person experiences.

Virtual try-on technology replaces a customer's wearing with arbitrary garments has significantly improved the online cloth shopping experience.
Thus, many fashion retailers devote their effort to relevant services and technologies, including ASOS's virtual catwalk app that allows a customer to see a cloth worn by a generic model in augmented reality~\footnote{https://www.retailgazette.co.uk/blog/2019/07/hands-asos-virtual-catwalk/}, and ZOZO's measurement suit that automatically measures the customer's body feature (e.g., hips size)\footnote{https://corp.zozo.com/en/news/20180501-3849/}.
At the same time, with the rise of deep image synthesis's quality, there arises a new urge for various image-based virtual try-on methods~\cite{han2018viton,wang2018toward,Yang_2020_CVPR}.
These methods seamlessly transfer a target garment in a product image to the corresponding region of a clothed person in a 2D image without resorting to 3D information.
Although these methods reduce the substantial labor costs and expert knowledge of existing 3D-based virtual try-on methods~\cite{shape_under_cloth:CVPR17,lahner2018deepwrinkles}, it inherently lacks the information required to generate accurately sized virtual results on a new body.
In addition, it limits the ability to react to possible interactions during the virtual try-on.
More specifically, it is difficult to synthesize the change of wrinkles caused by pose and body size change, as well as pulling and stretching of the garment by hand.
\chl{
The lack of simulated wrinkles prevents the user to understand how a garment looks on his/her body and to feel the physical properties of the material.
}
Because of these limitations, the customers fail to mimic the common practices they usually perform during physical try-on in the store.

In this work, we present an alternative workflow that handles rich interactions during real-time virtual try-on.
Our workflow is composed of two parts, \ie~garment capturing for the retailers to capture the rich deformations of the target garment, and photo-realistic clothed person image synthesis for the customers to experience the try-on result.
Distinguished from previous image-based virtual try-on works, the retailers capture each garment separately and train a synthesis model for each garment to provide rich visual feedback to the customers during try-on.
The customers only need to wear a measurement garment (\figname~\ref{fig:teaser}) to virtually trying-on various garments the retailers have captured.
The measurement garment helps our system segment and track the upper cloth region and other body regions accurately, thus supporting possible interactions compared to previous works.

For the retailers, we custom designed an actuated mannequin that can transform into diverse body sizes and poses through servo motors on upper arms and four adjustable regions in the torso.
All the degree-of-freedoms can be controlled independently, thus simulated wide ranges of possible body configurations.
The retailer only needs to put the target garment on the mannequin; our system automatically captures the dynamic appearance of the garment under a wide variety of poses, body sizes, and viewing angles. 

Our photo-realistic clothed person image synthesis part has two core components, \ie~the measurement garment design, and image-to-image mapping learning.
Existing image-based virtual try-on methods use a spectrum of body representations to generate virtual try-on results, including coarse body shape, 2D human pose, and semantic segmentation to predict the appearance of the transferred cloth and the remaining body parts.
However, these representations are often ambiguous and inaccurate to capture most common interaction during try-on.
Moreover, the estimations of these representations are often too slow for real-time virtual try-on. 
We proposed to use a measurement garment as a highly trackable body representation that enables real-time virtual try-on.
Inspired by~\cite{scholz2005garment}, we designed several patterns of measurement garments that are easy to be separated from the customers' body and can convey the customers' body configurations accurately.
Combined with the previously captured target cloth data, we learn the mapping between the measurement garment and each target cloth separately.
The mapping is learned using pix2pixHD model~\cite{wang2018pix2pixHD}, which replaces the image of the measurement garment to a target cloth.
During the interaction, a customer wears the measurement garment in front of a camera, and our system replaces the measurement garment with the target cloth on the screen.

We demonstrate the effectiveness of our proposed system using several target clothes and compared the synthesized results with a state-of-the-art virtual try-on method~\cite{Yang_2020_CVPR}.
We also conduct extensive ablation studies and a user study.
The purpose of the user study is to evaluate how helpful our virtual try-on system is for the customers to experience how a cloth fits to make an accurate buying decision.
And the study results suggested that the users are satisfied with the virtual try-on result and would love to use it while doing online shopping for clothes.

A possible concern is that it is too costly to manually
put each garment on the mannequin and capture many images spending hours \chl{thus limit the scalability of our method.}
It is true that it is much more work compared to try-on image synthesis from a single image.
Our current pipeline captures a single piece of garment in 2 hours,
therefore capturing a garment with 5 size variants sequentially requires
10 hours. This overhead is certainly a disadvantage of our method.
However, we believe that our method still has
commercial viability for two reasons.
First, the quality of resulting images are much higher than images synthesized from a single image. 
A single image simply does not provide sufficient information to synthesize realistic results
and it is necessary to take many images to achieve sufficient quality.
\chl{
Second, the person only needs to change the target garment on the mannequin and the rest of the process is automated. 
To further accelerate the capturing process, the retailers can prepare several mannequin-and-camera setups and capture multiple garments at the same time, and the mapping function can be trained in parallel using multiple GPUs.
}
Overall, we believe that the cost can be justified to provide high-quality virtual try-on experience and our method has reasonable scalability.

To summarize, the key contributions of this paper are 
\begin{itemize}
    \item We present a per garment capture and synthesis workflow for real-time virtual try on combing an actuated mannequin and measurement garment. It enables richer interactions with the garment compared to generic model trained with many garments.
    \item we present a prototype implementation with a custom-designed actuated mannequin and measurement garment. 
    The result shows  that synthesized try-on image appropriately respond to changes in body poses and sizes, as well as interaction such as pulling.
    \item we collected the first cloth dataset captured using our system covering ten target clothes under different body configurations that can benefit more virtual try-on research in the future.
\end{itemize}

\section{Related work}
\subsection{Virtual try-on}
The virtual try-on system~\cite{cordier2001from} aims to synthesize an image of a customer virtually wearing a fashion product such as clothing, eyeglasses, shoes, and makeup without physically wearing it.
With the widespread of online shopping, the virtual try-on system has become a crucial technology. 
However, it remains challenging to synthesize compelling images in real-time as the customers change poses; thus fail to provide a satisfied experience that can mimic their physical experience.
In recent years, the development of virtual try-on can be categorized into two kinds of methods: image-based method and 3D model-based method.
\paragraph{3D model-based virtual try-on}
The important ingredient of the 3D model-based virtual try-on system is to capture and reconstruct both the customers' body shapes and the garment geometry to replace the original garment with a new one. 
The garment modeling and capturing can be achieved by a RGB camera~\cite{xu20193d}, an additional depth camera~\cite{chen:2015:garment}, and a high-resolution 4D scanner~\cite{pons2017clothcap}.
On the other hand, to estimate the body shapes, many previous works perform body parts segmentation~\cite{gong2018instance,ruan2019devil,Gong_2019_CVPR} first, and then fit parametric body shape models (\eg~SMPL~\cite{loper2015smpl} or SCAPE~\cite{SCAPE}) to the person in the input photo~\cite{yang2018physics,xu20193d} or a sequence
of 3D point clouds~\cite{yang2016estimation}.
The captured data can be used to generate different garment detailed deformations~\cite{lahner2018deepwrinkles, CAPE:CVPR:20} and to map garment images directly onto a 3D human model~\cite{mir20pix2surf}.

One bottleneck of real-time virtual try-on is that high-quality garment simulation remains computationally very expensive.
In recent years, many researches aim for accelerating the garment simulation for better virtual try-on experiences using up-sampling techniques~\cite{example2010wang,kavan2011upsampling}, data-driven cloth simulation~\cite{near2013kim}, GPU-based simulation~\cite{Su2018gpu}, and machine learning approach~\cite{Santesteban2019LearningBasedAO}.

\paragraph{Image-based virtual try-on}
Another research direction that draws much attention is to achieve virtual try-on purely based on a 2D image without resorting to any 3D information.
Previous works leverage additional sensors (such as depth camera and infrared camera) to estimate the body shapes of the person~\cite{sekine2014virtual} or the fine wrinkles on the garment~\cite{avots2016automatic}.
The try-on results are generated by deforming the original garment~\cite{zhou2012image-based, yamada2014image} or image-based rendering~\cite{hauswiesner2013virtual} following the person's pose.
In more recent years, with the breakthrough in the deep generative model enables synthesizing plausible images including human faces~\cite{Karras2019stylegan2} and various objects~\cite{brock2018large}, many works aim to synthesize realistic clothed person in a simpler setting~\cite{lassner2017generative,jetchev2017conditional,raj2018swapnet, han2018viton,wang2018toward,Yang_2020_CVPR}.
These methods usually leverage body part segmentation generated by human parsing methods~\cite{gong2018instance,ruan2019devil,Gong_2019_CVPR} and body pose prediction~\cite{kokkinos2018densepose}.
These body representations are used to generated virtual try-on results with the same pose~\cite{han2018viton,wang2018toward,Yang_2020_CVPR} or the different pose~\cite{Dong_2019_ICCV, detail2019}.
One important and straightforward extension of these image-based virtual try-on method is considering the temporal coherence and provide video-based virtual try-on experiences~\cite{Dong_2019_ICCV,kuppashineon}.

While these methods produce plausible results, they share several limitations that prevent them in supporting common interactions during virtual try-on,
including (i) distorting the texture on the garment, (ii) blurring the body parts so that the identity is not well preserved.
Unlike these methods, we designed a measurement garment that is inspired by previous cloth capturing work~\cite{scholz2005garment, scholz2006texture}.
The measurement garment served as a fast trackable body representation that overcame the limitations of the common body representation used in previous works.

Moreover, most of the previous methods do not synthesize try-on results that correctly reflect fittingness, \ie~the relationship of the body shape of the person and the garment that is being worn.
Unlike previous methods, we propose a per garment capture and synthesis method, thus we can provide accurate fittingness for the customers.
We verified this by demonstrate results with different garment sizes and the user study by asking the participants to choose the correct size to buy while using the virtual try-on system. 
And all the participants choose the correct size for them.

\begin{figure*}[t!]
    \centering
    \includegraphics[width=\linewidth]{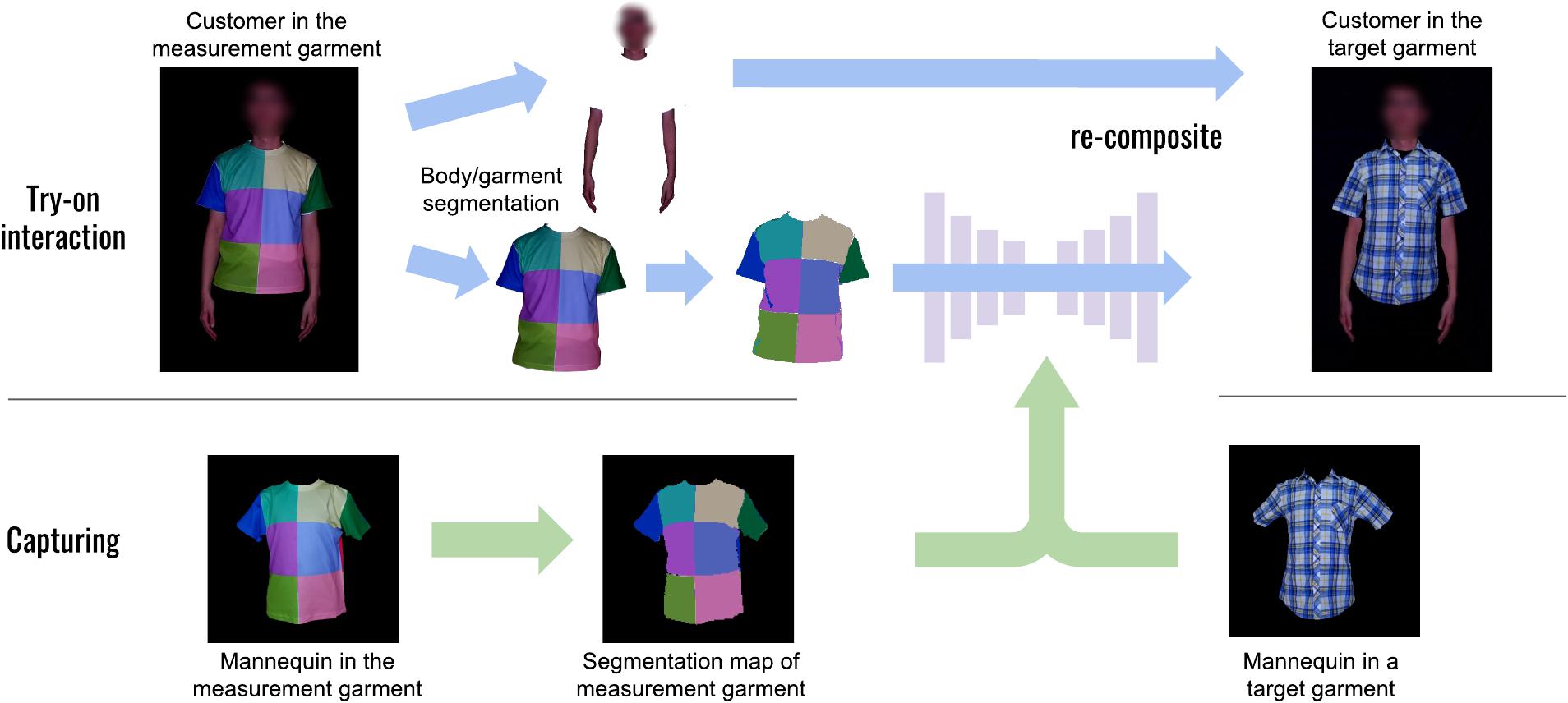}
    \caption{
        Overview of our system. 
        The retailer captured a set of images of the measurement garment, and a set of images for each target garment in the capturing studio. 
        Then they trained a image-to-image translation network for each target garment using the measurement garments' segmentation map and the target garment pair. 
        When a customer starts to try-on, they wear the measurement garment and our system provides high-fidelity virtual try-on for target garments that the retailer captured.
        And the user can perform common interactions with the garment, such as pulling and stretching the garment with hands.
    }
    \label{fig:overview}
\end{figure*}

\subsection{Cloth dataset}
In recent years, there are various fashion datasets contain thousands of clothes that support online shopping, personalized recommendation, and virtual try-on~\cite{cvpr16DeepFashion,DeepFashion2,zhu2020deepfashion3d,tiwari20sizer}.
Moreover, previous works~\cite{lahner2018deepwrinkles,pons2017clothcap} used specialized 4D capturing devices to capture the motions and wrinkles of a target garment worn by a person.
Another direction to quickly explore the detail deformations of a target garment on different body shapes is to use form-changing robotic mannequins~\cite{guo2016design,abels2013construction} that supports different body shapes~\cite{euveka,idummy,fitsme} and poses~\cite{palette}.
In our work, we use a custom-designed form-changing mannequin to capture the local and global deformations of the garment as our training data.
This mannequin brings several advantages, including (i) reduce the human labor costs for the retailers and (ii) enable an easy-to-control capturing process since we can change the body shapes and poses programmably.
\section{Overview}

To achieve realistic real-time virtual try-on and to support common interactions during try-on, we propose a per garment capturing module and a clothed person image synthesis module.
The garment capturing module captures RGBD images of the actuated mannequin wearing a garment (including the measurement garment and the target garments).
It produces the paired data between the measurement and the target garment under the same body sizes and poses.
The clothed person synthesis module then learns the mapping between the captured paired data using an image-to-image network.

To use our system, the retailer first needs to capture the measurement garment using the actuated mannequin in a capturing studio.
After capturing the measurement garment, the retailer can capture arbitrary target garments by repeating the same capturing process.
And then, they can use the captured paired data to train a image-to-image translation network for each target garment separately.
\myhl{
Existing workflow in the fashion industry is usually costly and time-consuming. 
It requires hiring professional models, photographers, and image editors and usually takes 3-5 days for the entire garment photograph capturing process.
Compared to the existing workflow, our proposed workflow can be cheaper because we do not need to hire costly professionals and can be faster by running the proposed workflow in parallel, \ie~the retailers can capture multiple garments using multiple actuated mannequins and train several image-to-image translation networks at the same time.
Most importantly, our workflow can provide a detailed virtual try-on experience with a customer's own body compared to the traditional static image of garments on a generic model.
}

When a customer comes into the fitting room, the customer first wears the measurement garment and stands in front of the video camera. 
This system does not require retraining the neural network for each customer. 
The customer first chooses a target garment among those captured by the retailer.
The customer can move in front of the camera and change poses just like what he/she will do during physical try-on. 


\begin{figure*}[t!]
    \centering
    \includegraphics[width=\linewidth]{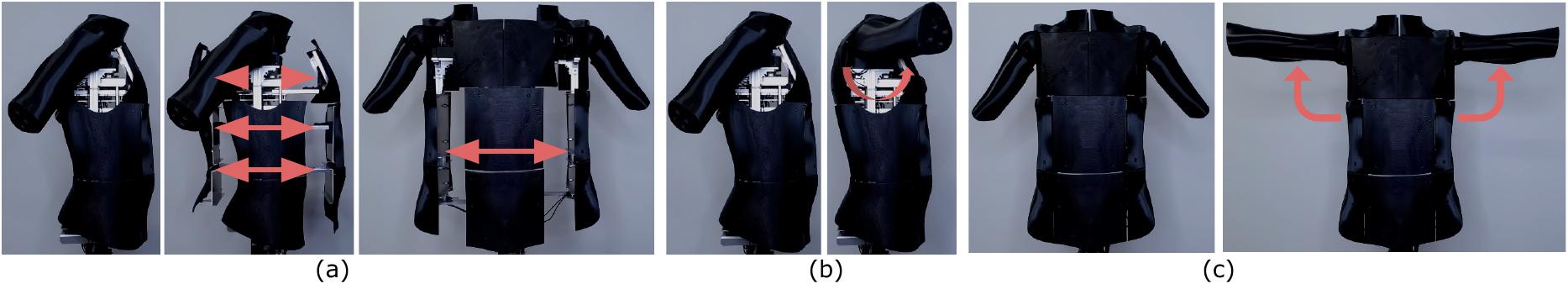}
    \caption{
Actuated mannequin degree-of-freedoms.
(a) The torso of our actuated mannequin is segmented into four adjustable regions (illustrated in red arrows), with three front-facing actuators and one side-facing actuator.
(b) (c) And each upper arm is actuated with two servo motors, allowing movement along the yaw and pitch axes.
}
\label{fig:man_dofs}
\end{figure*}
\section{Garment capturing}\label{sec:data_collection}
\subsection{Capturing device - Actuated Mannequin}
We designed the basic body shape of our actuated mannequin that approximates the average male body shapes in a commonly used parametric body shape model SMPL~\cite{loper2015smpl}.
To enable the mannequin to pose for a large variety of motions commonly found during physical try-out sessions in a physical store, we equipped it with nine degrees-of-freedoms:
\begin{itemize}
    \item independent arm movement for both arms (DoFs: $4$)
    \item four adjustable body region (DoFs: $4$)
    \item rotation around the center (DoFs: $1$)
\end{itemize}

More specifically, the torso is segmented into four adjustable regions, with three front-facing actuators and one side-facing actuator (\figname~\ref{fig:man_dofs}(a)).
And each upper arm is actuated with two servo motors, allowing movement along the yaw and pitch axes (\figname~\ref{fig:man_dofs}(b) and (c)).
All motors can be controlled independently and connected to the host computer, enabling automatic capturing.
The mannequin's arms and outer shell are 3D printed which is easy for reproduction, and it is strengthened with sheet metal internally.
We plan to release the design so that the actuated mannequin can be reproduced.


The mannequin is placed in front of a single color background, and we used an RGBD camera (Microsoft Azure Kinect DK) with fixed focal length, shutter speed, and exposure setting during capturing and testing.
We used color-based segmentation for separating the garment and the body parts; hence we avoid using the color of the background that is similar to the one of the target garment.
During our capturing, we used a black background for all the target garments and the measurement garment. 
We also dress the mannequin in a full-body skin-tight suit in the same color as the background.
The reason is that the surfaces of the mannequin generate reflections inference with the color-based garment segmentation.
In addition, the arm modules are detachable to allow garment changing.

\subsection{Capturing process}
To generate the virtual try-on result for each target garment, the retailer must captures data for both the target garments and a measurement garment (we will describe the design of the measurement garment in \secname~\ref{subsec:measurment_design}). 
However, measurement garment data can be reused to pair up with any new target garment the retailer wants to capture.
Thus, the retailer only needs to capture the measurement garment only once.
The procedure for capturing the measurement garment and the target garments are identical.
To capture a piece of garment, we first dress the mannequin in a skin-tight suit and set up a black background screen.
Afterward, we dress up the mannequin in the garment and begin the capturing session. 

We capture the garment by rotating the mannequin in a fixed body configuration (i.e., body size and arm pose). 
With our current setup, we collect $1$ image each degree the mannequin is rotated, and we captured $\pm67^{\circ}$ from the front of the mannequin. 
We divided the arms' movement range into two $5x5$ grids uniformly, and we moved both arms uniformly sampled from the grids. 
We adjusted the size of the body by $1cm$ both horizontally and vertically as we finish capturing a rotation, cycling through a predefined limit (0~20cm). 
In total, we collected $84375$ RGBD images for each piece of garment.
The entire process is automated except the initial setup and takes approximately $2$ hours for each piece of garment. 

\subsection{Data post-processing}\label{subsec:post_processing}
After capturing each target garment, we segmented the garment out from the background and obtain a garment mask by thresholding the depth values first and further refining the mask through color thresholding in HSV color space.
To align a pair of both the measurement garment and a target garment
under identical viewing angle and body configuration, we first crop the measurement garment using the garment mask.
The mask of the measurement garment always maintains a fix-sized border at the top and bottom relative to the height of the garment mask. 
We apply identical cropping parameters to the target garment to generate a direct match. 
Finally, we resized the images to $512px\times512px$ for learning.


\section{Synthesis and Try-on}
\label{sec:method}
Figure~\ref{fig:overview} details the workflow of the synthesis system. 
Taking the images of measurement and target garment as training data, we train an image-to-image translation network for each measurement garment and target garment pair. 

\begin{center}
\captionsetup{type=figure}
 \includegraphics[width=\linewidth]{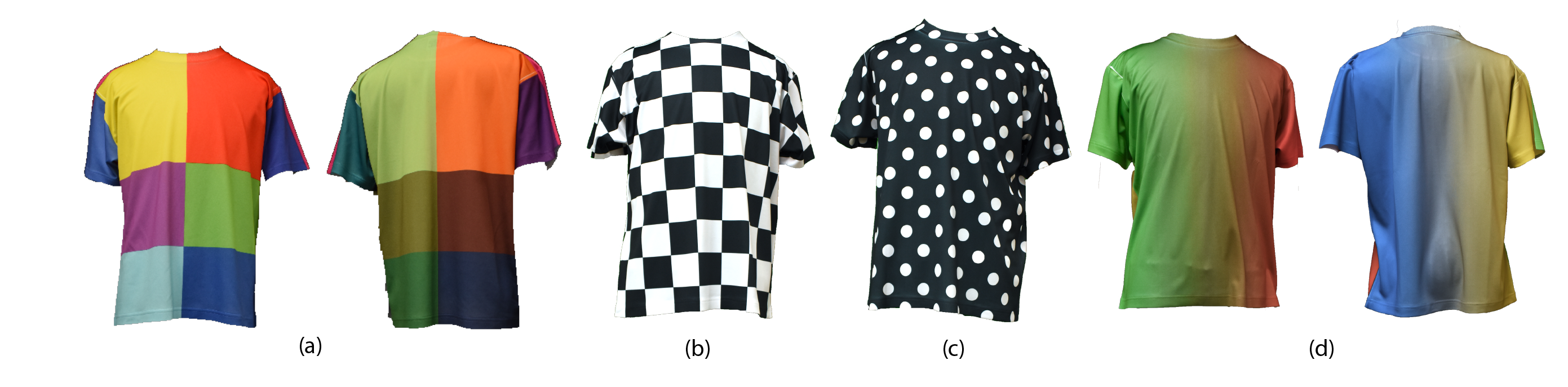}
 \caption{
We designed and tested measurement garments with (a) segmentation pattern, (b) checkerboard pattern, (c) dot pattern, (d) color gradient pattern. 
For (b) and (c) the front and back are identical.
 }
 \label{fig:tshirt}
\end{center}

\subsection{Measurement garment design}
\label{subsec:measurment_design}
One of the major differences between our system and many existing image-based virtual try-on systems is that we ask the user to wear a measurement garment during the interaction.
By defining the problem as an image-to-image translation from
measurement to target garment, we can capture rich garment deformations (\eg~wrinkles) of measurement garment during try-on and transfer it to the target garment.
Custom design measurement garment also makes image segmentation problem tractable, thus enabling real-time interaction.
Hence, there are several requirements for the design of the measurement garment. 
First, it should be separable from the customer's body (skin color) computationally efficient and robust under dramatic deformations.
Second, it should convey the body configuration (posture and size) of a customer. 
Third, it should allow the image-to-image translation network to easily translate it to a wide range of target garments without introducing additional artifacts. 
To address the above requirements, we designed and tested various patterns on the measurement garment including:
\begin{itemize}
    \item single color, no pattern (plain),
    \item color-coded, discrete segmentation (segmentation),
    \item color-coded, gradient image (gradient),
    \item monochrome, checkerboard pattern (checkerboard),
    \item monochrome, dot pattern (dot).
\end{itemize}
Please see figure~\ref{fig:tshirt} for a gallery of the measurement garments we designed and tested.
We concluded that the color-coded discrete segmentation garment is effective for synthesizing realistic and stable virtual try-on result, since the segmentation garment effectively divided the input image into multiple patches, thus allows the learning to perform more efficiently. 
In addition, by using a one-hot representation allows our garment extraction process to run at a much lower resolution, without compromising the image generation quality. 

\subsection{Image-to-image translation}
We learned the mapping between the measurement garment and each of the target garment using image-to-image translation technique.
The image-to-image translation function $f$ aims for mapping an image from input domain A ($I_A$) to output domain B ($I_B$): $I_B = f(I_A)$.
And it is widely used to translate images where input and output domains differ in surface appearances, \eg~from sketch or semantic labels to real-world photograph.
And the translation function $f$ is modeled as an conditional GAN~\cite{mirza2014conditional} where we trained coarse-to-fine generators and multi-scale discriminators followed Pix2PixHD~\cite{wang2018pix2pixHD}.
For a target garment ($G_t$), we use all the $84375$ captured images ($\mathbf{I}_{G_t}$) and the segmentation map of the measurement garment ($G_m$) captured image ($\mathbf{S}_{G_m}$) as training data. 
Given the paired training set $(\mathbf{S}_{G_m}, \mathbf{I}_{G_t})$, we seek a mapping funtion $f$ that satisfy: 
\begin{align}
M^i \odot {I}_{G_t}^{i} \approx f(M^i \odot {S}_{G_m}^{i}),
\end{align}
where $M^i$ is the garment mask of the $i$-th frame, ${S}_{G_m}^{i} \in \mathbf{S}_{G_m}$ is the $i$-th frame in measurement garment image set, and ${I}_{G_t}^{i} \in \mathbf{I}_{G_t}$ is the $i$-th frame in target garment image set.
We augmented the input image using standard image augmentation techniques, including random affine transform and minor color jittering. 
We applied a random translation of $\pm50px$ at each direction, random rotation within $\pm15^{\circ}$, shearing for $\pm10^{\circ}$. 
We also applied change of saturation, hue, brightness, and contrast for $0.05$. 
We used the default parameters provided by Pix2PixHD without hyperparameter tuning. 
We trained the system for ten epochs, which took approximately two days on a V100 GPU.
\subsection{Try-on}
During the interaction stage, a customer wearing the measurement garment and stand in front of our system. 
Generating the try-on result contains three stages: measurement garment extraction, target garment inference, and final composition (as shown in the upper row in \figname~\ref{fig:overview}). 

\paragraph{Measurement garment extraction}
The system extracts the garment from the input camera feed using a three-step process. 
A rough whole body mask is first generated using the depth information. 
We then applied skin detection to the image, which removes the exposed arm, face, and neck of the user and leaves only the garment in the image. 
The garment is then segmented into the eight color patches using K-Means clustering in HSV color space. 

\paragraph{Target garment inference}
After extracting the garment mask, we cropped the frame using the cropping parameters chosen at \secname~\ref{subsec:post_processing}. 
The cropped image is then masked with the garment mask and used as the input to the image-to-image translation model to infer the resulting target garment. 

\paragraph{Final re-composition}
The output result of the image-to-image translation model is then re-composed back into the original frame. 
We do so by first applying the post-processing (cropping and resizing) in reverse order. 
We automatically generated a garment mask from the network output using the intensity value to avoid overlay body regions that should not be covered.
Currently, we assume the garment should be always on top and should remain only one single component.
So if the arm stays on top of the garment that separate the garment into two parts, our method currently fails to map the target garment to the measurement garment.
\section{Results and evaluations}
To demonstrate the effectiveness of our system, we performed a series of experiments, including ablation studies, and compared our synthesized result with the state-of-the-art virtual try-on method~\cite{Yang_2020_CVPR}.
Furthermore, we discussed how our workflow better supports the possible interaction during virtual try-on.

\subsection{Implementation and performance}
Our setup consists of a desktop PC with a Ryzen 1700 3.4 GHz CPU, 32GB RAM, and a single GTX2080Ti. 
Our performance on the real-time system is approximately 10FPS. 

\subsection{Try-on result comparison}
We compare our method with a state-of-the-art image-based virtual try-on method~\cite{Yang_2020_CVPR}. 
To evaluate the try-on results' quality, we designed five preset motions that are commonly performed during upper body garment try-on.
The preset motions include
\begin{itemize}
    \item \textit{T-pose}: raising upper arms horizontally 
    \item \textit{pulling}: stretching the garment
    \item \textit{side}:  rotate the torso to both sides
    \item \textit{rolling}: raise upper arms and stretch forward and backward. 
    \item \textit{butterfly}: raise upper arms up and forward
\end{itemize}
Please check the detail of these motions in our supplemental videos.
We recruited 5 participants and asked them to perform these motions in the measurement garment and $4$ target garments.
We recorded all their motions as ground truth videos, and use them to evaluate the quality of different try-on results.

With the recorded ground truth motion videos,
we compared the following two virtual try-on methods:
\begin{itemize}
    \item Our system (\textit{Ours})
    \item \textit{ACGPN}, the state-of-the-art image-based virtual try-on system  \cite{Yang_2020_CVPR} (we used their public implementation\footnote{\url{https://github.com/switchablenorms/DeepFashion_Try_On}})
\end{itemize}
\textit{ACGPN}~\cite{Yang_2020_CVPR} warps the input garment image to fit the pose of the body using compact body representations including semantic segmentation and keypoint locations from pose estimation.
It synthesizes the try-on results with highest quality on both the body parts and retargetted garment compared to other virtual try-on methods~\cite{han2018viton,wang2018toward}.

\begin{figure}[!ht]
    \centering
    \includegraphics[width=\linewidth]{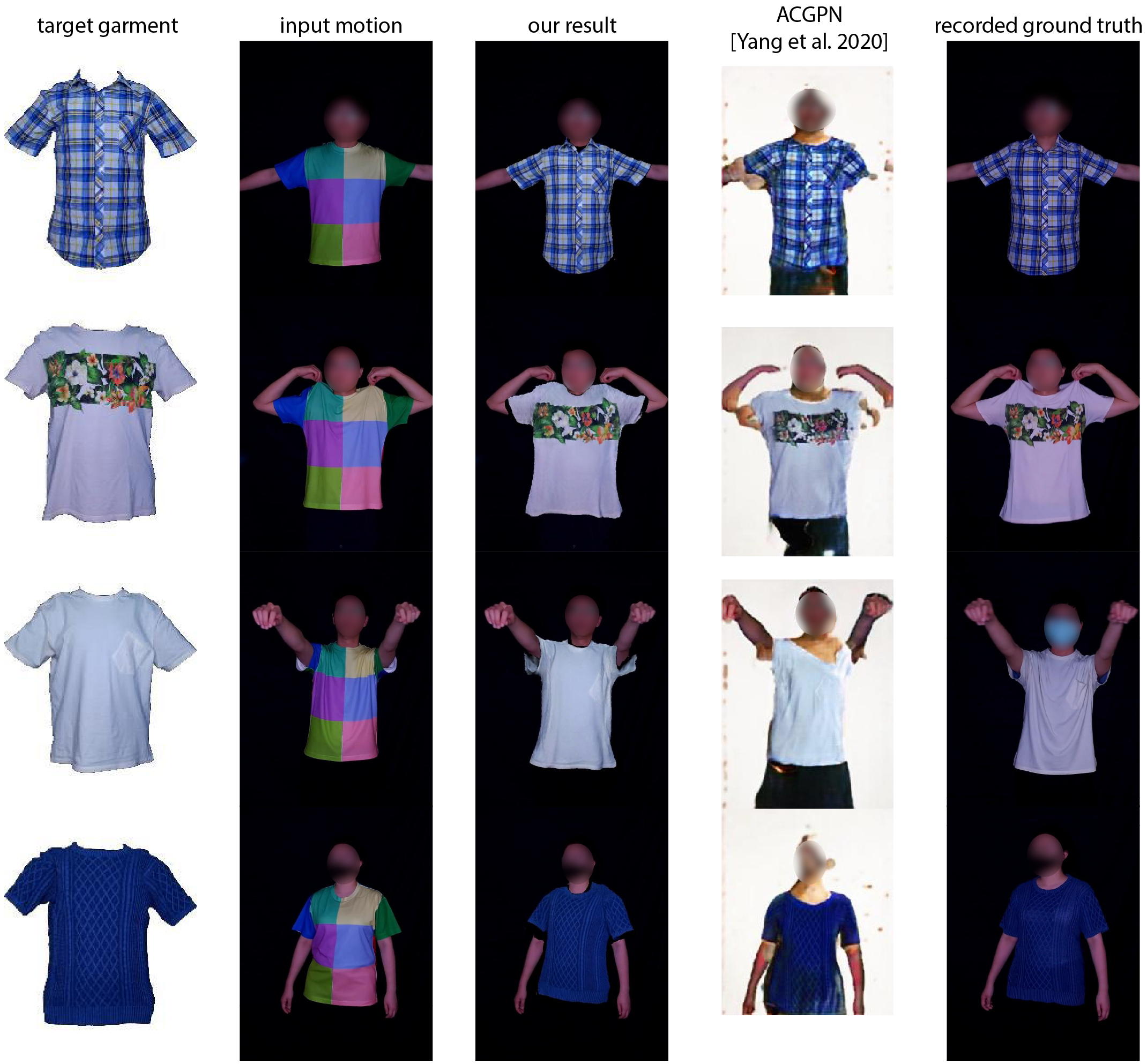}
  \caption{
  Our try-on results on various target garments.
  Given the frame of input motion, our method synthesizes results with clearer garment and body parts.
  We showed the ground truth in the last column for reference.
}
\label{fig:target_garments_result}
\end{figure}

The input to ACGPN~\cite{Yang_2020_CVPR} is the recorded motion videos where the participants wearing the measurement garment (as shown in the 2nd column of \figname~\ref{fig:target_garments_result}). 
We preprocessed (cropping and resizing) the input videos to 192x256 and then apply \textit{ACGPN} frame-by-frame without resorting to any temporal information.
Noted that our synthesis method does not leverage temporal information as well.
We show the comparison results of example frames in \figname~\ref{fig:target_garments_result} (please see the full video comparison in the supplemental video).
By comparing to the ground truth (last row in \figname~\ref{fig:target_garments_result}), we can observe that under the common try-on motions, our synthesis method can generate realistic try-on result. 
More specifically, our try-on results are with (i) detailed wrinkles, (ii) preserved textures on the garment, and (iii) clear body parts compared to the previous work~\cite{Yang_2020_CVPR}.
Moreover, our method synthesizes try-on results in higher resolution ($512x512$) than \textit{ACGPN} ($192x256$).

\chl{
We observed the quality of ACGPN results in \mbox{\figname~\ref{fig:target_garments_result}} are degradated from the results shown in the original paper.
The major reasons are pose variation and data preprocessing. 
In terms of pose variation, most of the existing virtual try-on works used \mbox{VITON dataset~\cite{han2018viton}} a collection of professional model pictures as training data, which only contains a limited variation of poses, while our participants moved freely during testing. 
Furthermore, the detailed preprocessing of the original images (scaled, cropped, and centered) used in ACGPN are unspecified in the paper. 
Thus, we tried to reproduce the preprocessing but it was difficult to fully replicate the results in the original paper.
}

\subsubsection{Try-on result discussion and ablation studies}
\begin{figure*}
\centering
  \includegraphics[width=\linewidth]{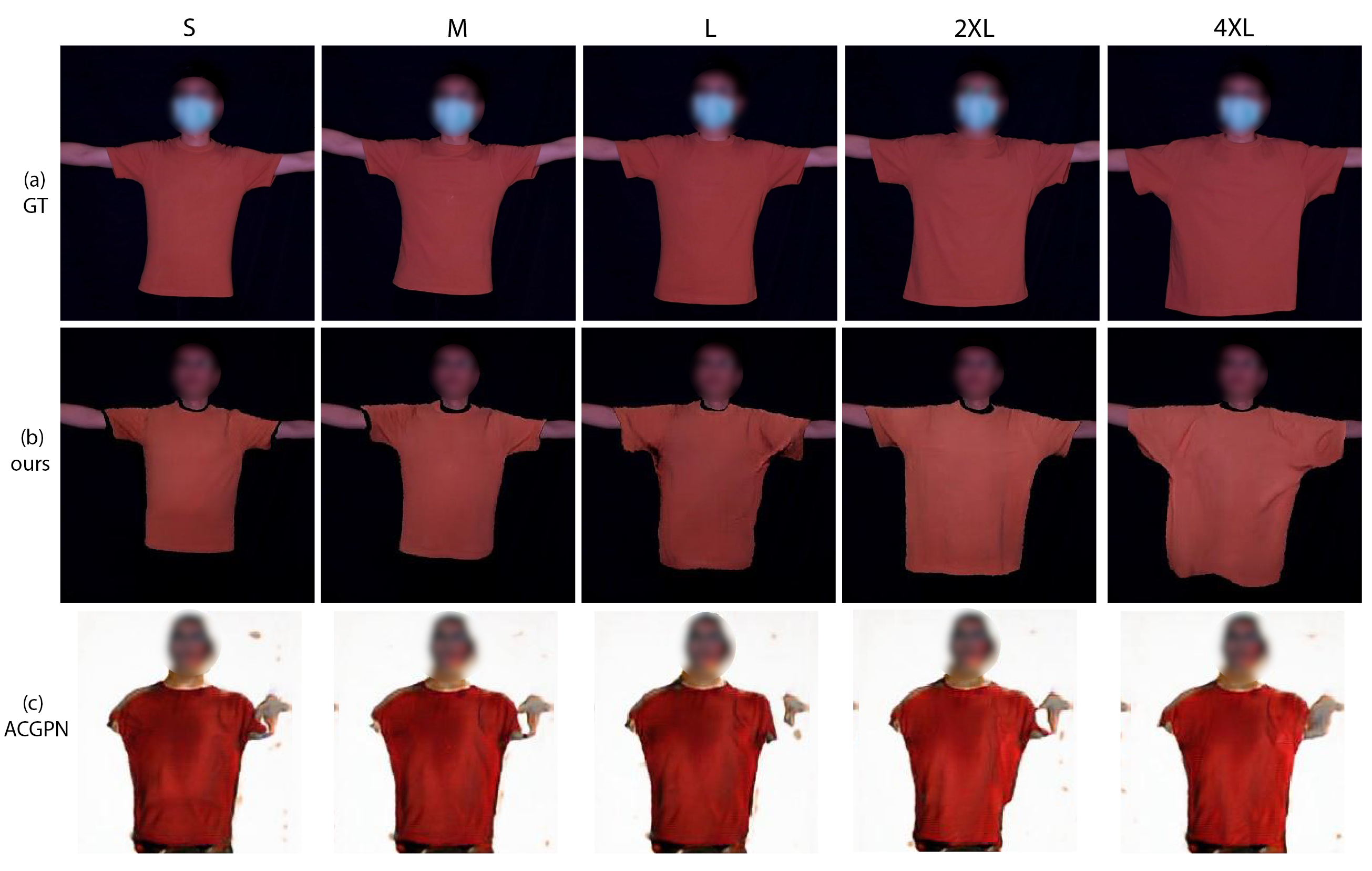}
    \caption{
    Diverse garment sizes.
    We capture (a) the same T-shirt design in different sizes as ground truth.
    (b) Our system can synthesize the try-on results for different sizes, generated using the same measurement garment. 
    (c) ACGPN~\cite{Yang_2020_CVPR} fails to synthesize results that convey different garment sizes.
    }
    \label{fig:different_size}
\end{figure*}
\paragraph{Diverse garment sizes}\label{sec:fit_robust}
To provide a realistic try-on experience, it is essential that the try-on system correctly depicts the target garment's fittingness worn by a customer. 
More specifically, our system can reflect the fittingness of different garment sizes.
We captured the same T-shirt design of the different sizes separately and trained individual networks for each garment size.
As shown in \figname~\ref{fig:different_size}, we can see that our system can synthesize the try-on results for different sizes with subtle details such as clothes slack.
Simultaneously, the existing deep image-based methods use a single product image as input fails to synthesize such variations and details.
In addition, we also showed in our user study (\secname~\ref{sec:user_study}) that the participants could accurately select the size that they preferred. 

\begin{figure}[t!]
\centering
  \includegraphics[width=\linewidth]{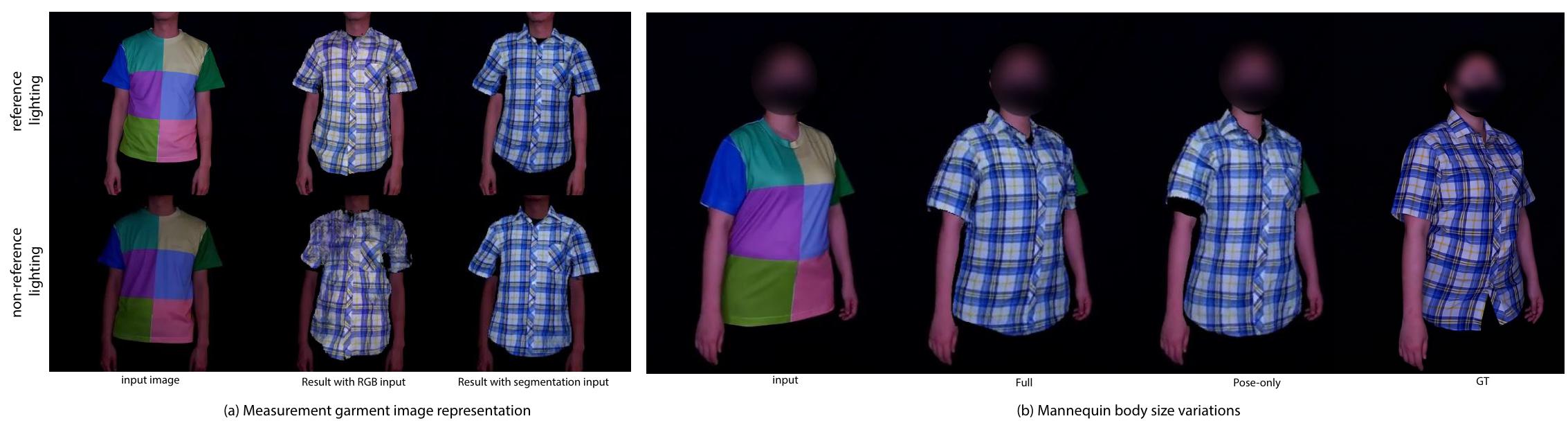}
  \caption{
  (a) Measurement garment image representation. With the reference lighting (the same lighting used during the capturing), our synthesis method can synthesize similar quality of try-on results with both RGB input and segmentation map input. 
  However, when we use lighting that is different from the reference lighting, the try-on result synthesized with RGB input becomes worse, especially fails to synthesize the pattern on the garment.
  (b) Mannequin body size variations.
  Given the input frame, the synthesis model trained with the \textit{Full} dataset synthesizes the try-on result better depicts the body shapes compared to the model trained with \textit{Pose-only} dataset.
We also show the ground truth (\ie~the same participant physically wearing the target garment) in the last column.
  }
\label{fig:repr_size}
\end{figure}
\paragraph{Measurement garment image representation}
We compared two input representations of the color-coded measurement garment: (i) RGB image and (ii) segmentation map of the measurement garment. 
The synthesized results under different lighting conditions using these two input representations are shown in \figname~\ref{fig:repr_size}(a).
We can observe that the network trained with the segmentation map is more robust under different lighting conditions; thus synthesize better try-on result.


\paragraph{Mannequin body size variations}
We compared the try-on results synthesized by two networks that use two sets of data
\begin{itemize}
\item \textit{Pose-only}: we captured this dataset with only pose variations (\ie~we did not use the four DoFs on the mannequin's torso to change the body size)
\item \textit{Full}: we captured this dataset with the full body poses and sizes variations.
\end{itemize}
We showed the results in \figname~\ref{fig:repr_size}(b), where we can observe that the network trained with the \textit{Full} dataset can synthesize the try-on result that depict body shape better and closer to the ground truth image.

\begin{figure}[t!]
    \centering
    \includegraphics[width=\linewidth]{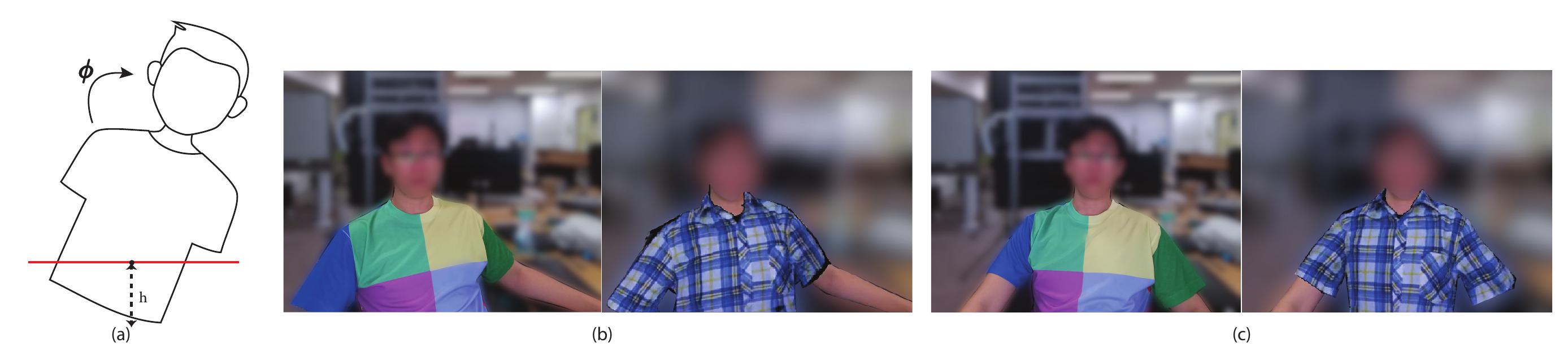}
  \caption{
  (a) Partial upper body data augmentation. We randomly rotate the upper body by $\phi$ degree and set the clipping line (the red line) at $h$.
  And we only use the image content above the clipping line as the new training data.
  (b)(c) Two example frames where we change the measurement garment into a shirt during a video conference.
}
\label{fig:zoom_demo}
\end{figure}

\subsection{Cloth changing for video conference}

In recent days, we often have professional remote video conferences with colleagues due to pandemic.
Our method can virtually change the upper body clothing without physically changing to formal clothing such as shirt.

\paragraph{Partial upper body data augmentation}
\myhl{
Unlike most of the results shown above, where the whole upper body is captured, only the top part of users' upper body is captured during video conferences.
To support the video conference scenario, we perform a data augmentation on the data captured using our mannequin.
For each garment we captured in \mbox{\secname~\ref{sec:data_collection}}, we randomly rotate the image by $\phi$ degree and create a clipping line that is parameterized by the height value $h$ (\mbox{\figname~\ref{fig:zoom_demo}(a))}.
And we use this clipping line to split both the measurement garment image and the target garment image in the training pair, and only use the image content above the clipping line.
}

\myhl{
We change the cloth of two example video conference frames using our method in \mbox{\figname~\ref{fig:zoom_demo}}(b) and (c).
Please see the accompanied video for more detail demonstration.
}

\section{User study}
\label{sec:user_study}
We conducted a user study with the goals to evaluate: (i) the importance of different aspects of a virtual try-on system and (ii) whether our system provides a convincing try-on experience to the customers. 
We designed our user study following~\cite{mixedVR2013} and included additional questions regarding concerns over our system. 
The participants first answered a questionnaire regarding their habits and preferences of online shopping for fashion items, as well as their general attitude towards virtual try-on.
After answering the questionnaire, the participants tested our system freely. 
Our virtual try-on result was displayed in front of the user in real-time. 
In total, we provided $10$ target garments for the participant to try-on.
After this free form try-on session, the participants answered a questionnaire regarding our system on whether it provide a convincing experience for shopping. 

\subsection{Participants}
We recruited $8$ participants ($5$ males-identifying and $3$ females-identifying) interested in purchasing clothing items online. 
Of the $8$ participants, $7$ have experienced purchasing fashion items online, $4$ participants purchase online at least once per season.

\subsection{User study procedure}
Before the user study, the participants are asked to fill in a pre-study questionnaire to collect demographic information and their online shopping impression and experiences. 
Then the participant wore the measurement garment and entered the virtual fitting area. They were given 3 minutes to move freely inside the area to practice different movements and get used to the system. During the practice time, the participant can see the raw camera feed without the virtual try-on effect. 
After the practice time, we introduced the virtual try-on system to the user. 
The participants were given 10 minutes to experience the system, freely moving in front of the camera with a piece of upper body garment they chose. 
We provide $10$ target garments to the participants to try-on.
We separate these garments into two sets:
\begin{itemize}
\item $5$ garments with different designs
\item $5$ garments with the same design but in different sizes (S, M, L, XXL, 4XL).
\end{itemize}
We randomized the order of the size of the garments being presented to the participants and did not tell participants the size of the virtual garment. 

To demonstrate that our system can provide correct fittingness during the virtual try-on, we asked the participants to identify one garment from $5$ different sizes that they think is the most suitable for them solely based on the virtual try-on process. 
To verify the selection, after the try-on session, we gave the participants all the five physical \textit{sized garments}, and we asked them to identify which garment is the most suitable for them physically.
Finally, the participants answered the final questionnaire regarding our system and whether it can help them make a more informed purchase decision without physically try-on the real garments.


\begin{figure}
\centering
  \includegraphics[width=\linewidth]{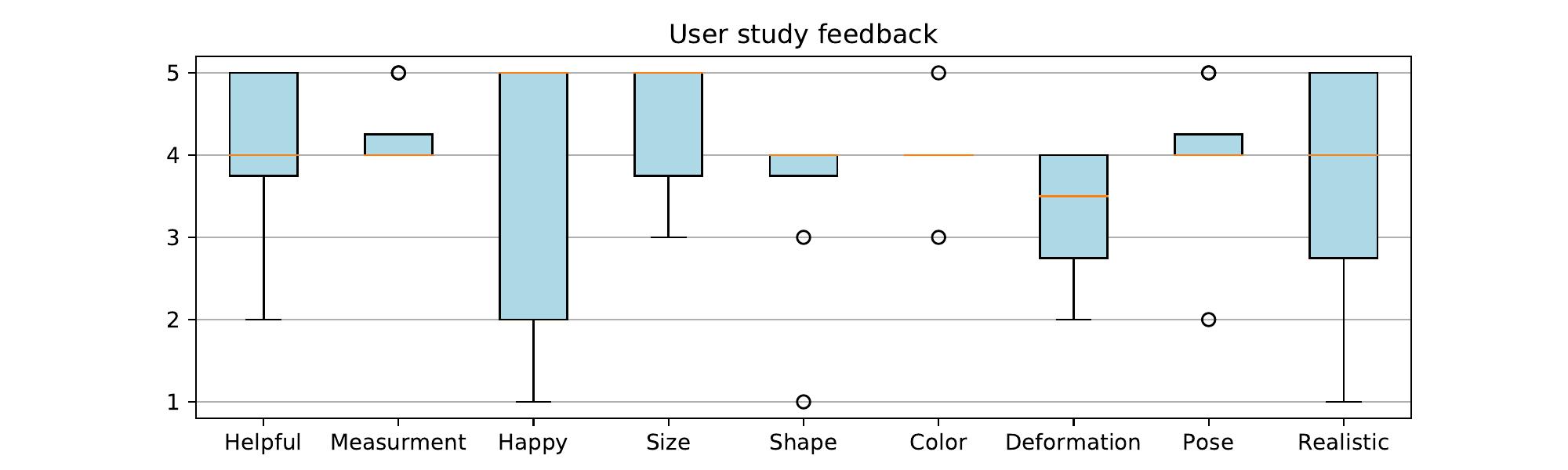}
    \caption{
    Box plots of the user study feedback.
    On the 5-point Likert scale, the participants rated our system on \textit{Helpful}: helpfulness on assiting them to make better purchase decisions ($4.00(STD=1.07)$), \textit{Measurement}: acceptability to wear the measurement garment ($4.25(STD=0.46)$), \textit{Happy}: whether they would be happy to use it ($3.75(STD=1.75)$), \textit{Size}: depicting the size ($4.38(STD=0.97)$),  \textit{Shape}: shape ($3.50(STD=1.07)$), \textit{Color}:color (e.g., logos and color) ($4.00(STD=0.53)$), \textit{Deformation}: deformation of the garment ($3.25(STD=0.89)$), and 
    \textit{Pose}: pose ($4.00(STD=0.93)$), respectively.
    Finally, the participants agreed that \textit{Realistic}: the garments looked realistic ($3.63(STD=1.50)$).
    }
    \label{fig:user_study_rating}
\end{figure}
\subsection{User study result}
\paragraph{Criteria for a useful virtual try-on system}
In our pre-study questionnaire, we asked our participants what is the important feature for the virtual try-on system in making their purchase decision on a 5-point Likert scale, with $1$ being the least important and $5$ being the most important. 
The participants considered the ability to accurately reflect the size of the garment relative to their bodies the most important $4.5 (STD=1.07)$. 
The ability to mix-and-match their own garment was also very important $4.38(STD=0.74)$. 
Accurately representing graphical design (e.g., logos) on the garment and being real-time is considered equally important with scores $4.00(STD=1.20)$ and $4.00(STD=1.07)$ respectively.
Viewing themselves with different poses and from different angles were relatively less important, rated at $3.88(STD=0.83)$ and $3.63(STD=1.30)$ respectively.
Finally, the participants are least concerned about the ability to view the garment under different lighting conditions $3.50(STD=1.07)$. 

\paragraph{Feedback on our system}
The participants provided feedback on our system, as summarized in \figname~\ref{fig:user_study_rating}.
On the 5-point Likert scale, the participants agreed that our system helped them make better purchase decisions in general ($4.00(STD=1.07)$), and depicted the garment size an color accurately.
Furthermore, they agreed that our system respond to their pose changing appropriately.
Finally, \chl{the participants consider the quality of garment is sufficient for virtual try-on} ($3.63(STD=1.50)$).

For the fittingness test, all except one participant correctly identified the correct size that they would like to purchase using only our system (\ie~the garment they chose using our system and the garment they chose after trying them are identical), even though the preferred size varies largely among the participants \myhl{(2 participant preferred S, 3 participants preferred M, 2 participant preferred L, and 1 participant preferred XXL)}.

\section{Limitation and Discussion}
\begin{figure}
\centering
  \includegraphics[width=\linewidth]{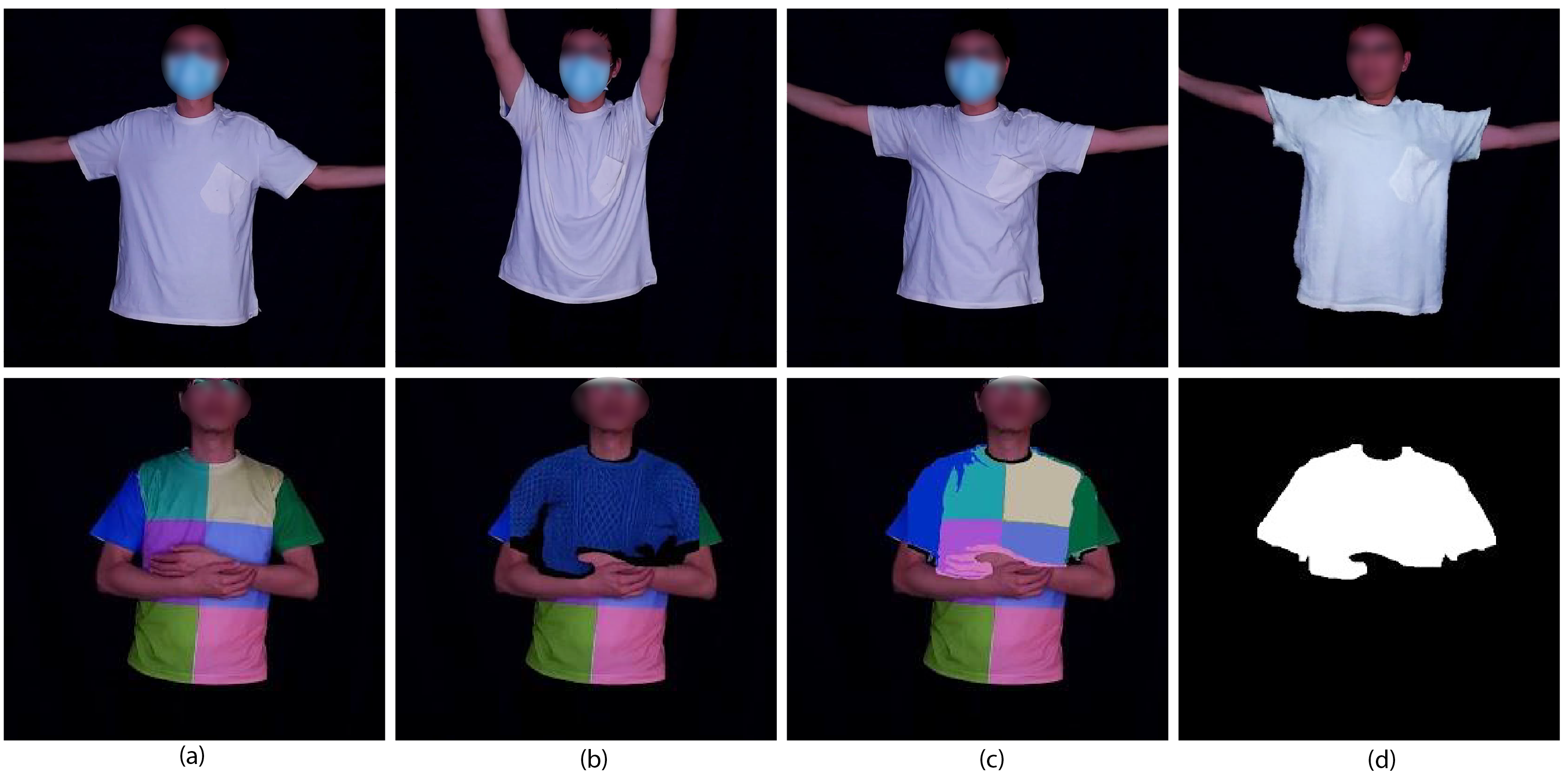}
    \caption{
    Limitations and failure cases.
    Top: (a)-(c) captured sequence of a participant raises both arms and puts them down. 
    (d) Since our method only uses single frame information as input, we can not synthesize the garment slacks caused by a sequence of motion 
    (the pose in (c) and (d) are the same).
    Bottom: (a) given the input image, (b) our method can not synthesize the full garment due to the arm occlusion.
    We also show (c) the segmentation map and (d) the garment mask of the input image as references. 
    }
    \label{fig:limitation}
\end{figure}

\paragraph{Utilize temporal information}
Our system uses single frame measurement garment segmentation map as input. 
As a result, it is incapable of generating deformation generated from sequence of movement.
For example, after a sequence of motion (\figname~\ref{fig:limitation}(a)-(c)), there are many wrinkles on the captured garment.
However, the result synthesized by our method (\figname~\ref{fig:limitation}(d)) fails to reproduce those wrinkles.
\paragraph{Structurally different garment}
Our system uses a measurement garment to predicts the shape and texture of the target garment.
When the structure of the target garment is drastically different from that of the measurement garment or partially occluded by arms (see bottom row of \figname~\ref{fig:limitation}), our method fails to synthesize high fidelity results.
For example, with the short-sleeved measurement garment, our system fails on the long-sleeved target garment since the target garment's information at the forearm position is not captured.

\paragraph{Advanced mannequin}
We plan to develop controllable forearms and torso that achieve an even wider variety of body pose changes to mimic more diverse human try-on motions and support more garment structures.
\chl{For example, we can design a full-length arm for the mannequin in order to support long sleeves garments.}

Moreover, we designed our current mannequin by referring to the male body shape.
We plan to design detachable components which allow capturing garments in female body shape.
\paragraph{Capture more detailed garment information}
In the future, we plan to add more cameras to take images from more diverse viewing positions, such as positions from different vertical angles. 
We also plan to control lighting conditions such that our dataset contains additional information for the algorithms to infer the deformations.
Also, we plan to perturb the garments by shaking them or blowing air to them to collect more potential deformations.
\paragraph{Advanced measurement garment design}
In our work, we proposed to use measurement garment as a better body representation for virtual try-on compared to other body representations used in previous works (\eg~body parts segmentation label).
We plan to investigate other measurement garment designs, \chl{\eg~more complicated color patterns used in\mbox{~\cite{White:2007:CAO,wang2011practical}}} or besides the color-coded measurement garment.
For example, we would like to optimize a measurement garment given a set of target garments.
The potential workflow is that retailers can release a new measurement garment together with a set of newly released target garments every season.

\section{Conclusion}
We proposed a per garment capture and synthesis workflow to support real-time virtual try-on.
Our key contributions are the use of actuated mannequin for capturing garment geometries and the introduction of the measurement garment for fast and accurate garment tracking that supports virtual try-on using the common interactions.
Our per garment workflow captures extensive amount of images of the target garments, thus handling rich common try-on interactions, including pulling or stretching the garments.
Moreover, it enables the users to accurately depicting the fittingness of a target garment.
In our user study, the participants are satisfied with our method for providing a realistic try-on experiences and adapting to different body sizes and poses.
Finally, our system also have a potential application for virtual costumes in video conferences. 
For user who is working from home, they can wear a light and comfortable measurement garment while looking professional to the other side of the screen.




\section{Acknowledgement}
We warmly thank all the participants for their time and comments regarding our system. 
We also thank the anonymous reviewers who helped us improve the paper. 
We especially thank Dr. Reo Matsumura, CEO of karakuri products Inc., for collaborating with us on designing and manufacturing the mannequin robot.
This work was supported by JST CREST Grant Number JPMJCR17A1, Japan.
I-Chao Shen was supported by JSPS KAKENHI Grant Number JP21F20075.

\bibliographystyle{ACM-Reference-Format}
\bibliography{mannequin}
\end{document}